\documentclass[aps,prb,twocolumn,superscriptaddress,showpacs,amsmath,amssymb,longbibliography, amsfonts,10pt]{revtex4-2}
\usepackage{graphicx} 
\usepackage{amsmath}
\usepackage{amsfonts}
\usepackage{amssymb}
\usepackage{graphicx}
\usepackage{color}
\usepackage{xcolor}
\usepackage{bbold}
\usepackage{appendix}
\usepackage[citecolor=blue]{hyperref}
\hypersetup{colorlinks}
\usepackage{subcaption}
\usepackage{bookmark}
\usepackage{makecell}
\usepackage{mathtools}
\usepackage{comment}
\usepackage{tikz}
\usetikzlibrary{arrows.meta}

\usepackage{array}
\usepackage{multirow}
\usepackage[nolist,nohyperlinks]{acronym}
\usepackage[export]{adjustbox}

\newcommand{\vecq}{\boldsymbol{q}}
\newcommand{\xx}{\boldsymbol{x}}

\newcommand{\vecphi}{\boldsymbol{\phi}}
\newcommand{\dd}{\mathrm{d}}
\newcommand{\nbath}{n_{\mathrm{b}}}

\definecolor{rotatingColor}{rgb}{0.741176, 0.811765, 1}
\definecolor{xyColor}{rgb}{0.611765, 0.905882, 0.658824}
\definecolor{paraColor}{rgb}{0.952941, 0.952941, 0.913725}
\definecolor{cepColor}{rgb}{0.858824, 0, 0}

\begin{document}

\title{Nonequilibrium orders in parametrically driven field theories}

\author{Carl Philipp Zelle}\affiliation{Department of Physics, Harvard University, Cambridge MA 02138, USA}
\author{Romain Daviet}\affiliation{Institut f\"ur Theoretische Physik, Universit\"at zu K\"oln, 50937 Cologne, Germany}
\author{Andrew J. Millis}\affiliation{Center for Computational Quantum Physics, The Flatiron Institute, 162 5th Avenue,New York, NY 10010}\affiliation{Department of Physics, Columbia University, 538 West 120th Street, New York, New York 10027}
\author{Sebastian Diehl}\affiliation{Institut f\"ur Theoretische Physik, Universit\"at zu K\"oln, 50937 Cologne, Germany}

\begin{abstract}
    Driving quantum materials with coherent light has proven a powerful platform to realize a plethora of interesting phases and transitions, ranging from ferroelectricity to superconductivity and limit cycles in pumped magnonics. In this paper we develop the field theoretical framework to describe nonequilibrium phases that emerge in systems pumped by rapid parametric drives. We consider paradigmatic $O(N)$ models that describe the long-wavelength fluctuations of ordering fields in many condensed matter set ups. We show that rapid parametric driving of these models can induce an effective pump mechanism in the long wavelength regime through nonlinear scattering. This induces a nonequilibrium transition into a time-crystalline phase.
\end{abstract}

\date{\today}
\maketitle

\section{Introduction}
The classification and description of nonequilibrium phases of matter as well as their realizations in physical systems are major research endeavors in modern  physics~\cite{Taeuber2014}. Although the mechanisms of phase transitions and their universal behaviors set by dimension and symmetry are (in many cases) well understood in thermal equilibrium \cite{ZinnJustin}, the situation is much less clear for nonequilibrium steady-states.\\
One of the most prominent resources of nonequilibrium conditions in recent experiments is laser pumping of a material coupled to a bath. In various setups, resonant driving of a high frequency mode can stabilize a macroscopic nonequilibrium steady-state at low frequencies. Examples range from driven-dissipative Bose condensation of exciton-polaritons in cavity pumped materials \cite{imamoglu1996,kasprzak2006,balili2007,deng2002,Fontaine2021,byrnes2014,deng2010,Carusotto2013} to transient light induced ferroelectricity \cite{nova2019,li2019ferro} and superconductivity at temperatures well above the equilibrium superconducting critical temperature \cite{fausti2011,mitrano2016,cremin2019,buzzi2020,fava2024}. Furthermore, pumping magnetic structures leading to driven nonlinear magnonics \cite{zhang2024,Shan2024,kaplan2025} and magnon condensates observed in laser driven Yttrium-iron-garnet (YIG) films \cite{Demokritov2006, Demidov2007, NowikBoltyk2012} represents a fruitful avenue to create nonthermal phases of matter.
\\
In this paper we show that rapid parametric drives, typically used to model such laser-driven systems \cite{Chandran2016,kennes2017, eckhardt2024, Walldorf19, Diessel2026,okugawa2026, Hosseinabadi2026}, can induce an antidamping instability in order parameter field theories. This arises through scattering of high frequency modes created by the drive into the low frequency regime. This constitutes an effective incoherent single particle pump, or antidamping, much akin to the mechanism of lasing in cavity exciton-polaritons \cite{Carusotto2013}. This antidamping instability gives rise to a dynamic long range order, tracing out a limit cycle \cite{zelle2024}, which is evidently nonthermal \cite{Bruno2013, Watanabe2015, daviet2024, daviet2025}.\\
While these previous works have assumed antidamping to emerge on phenomenological grounds, here we provide an explicit yet generically applicable derivation of this effect based on nonequilibrium diagrammatics. A key takeaway is that one needs to take effective two-loop effects, that are next to leading order in large $N$ \cite{Chandran2016, Diessel2026}, into account. The relevant diagrams are those which, at equilibrium, are responsible for thermalization — they describe scattering processes which can redistribute modes in frequency and momentum space. Under conditions of parametric drive, which creates high frequency and momentum modes, they open a channel to incoherently pump and also heat long wavelength degrees of freedom. We provide generic estimates for this competition based on an analytic evaluation of the respective two-loop diagrams in perturbation theory.
 The emergence of a slow limit-cycle order is confirmed through direct numerical simulations of the underlying dynamics.

\begin{figure*}
     \centering
       \subfloat[\label{fig:phase_diagram_a}]{%
  \includegraphics[width=0.45\textwidth]{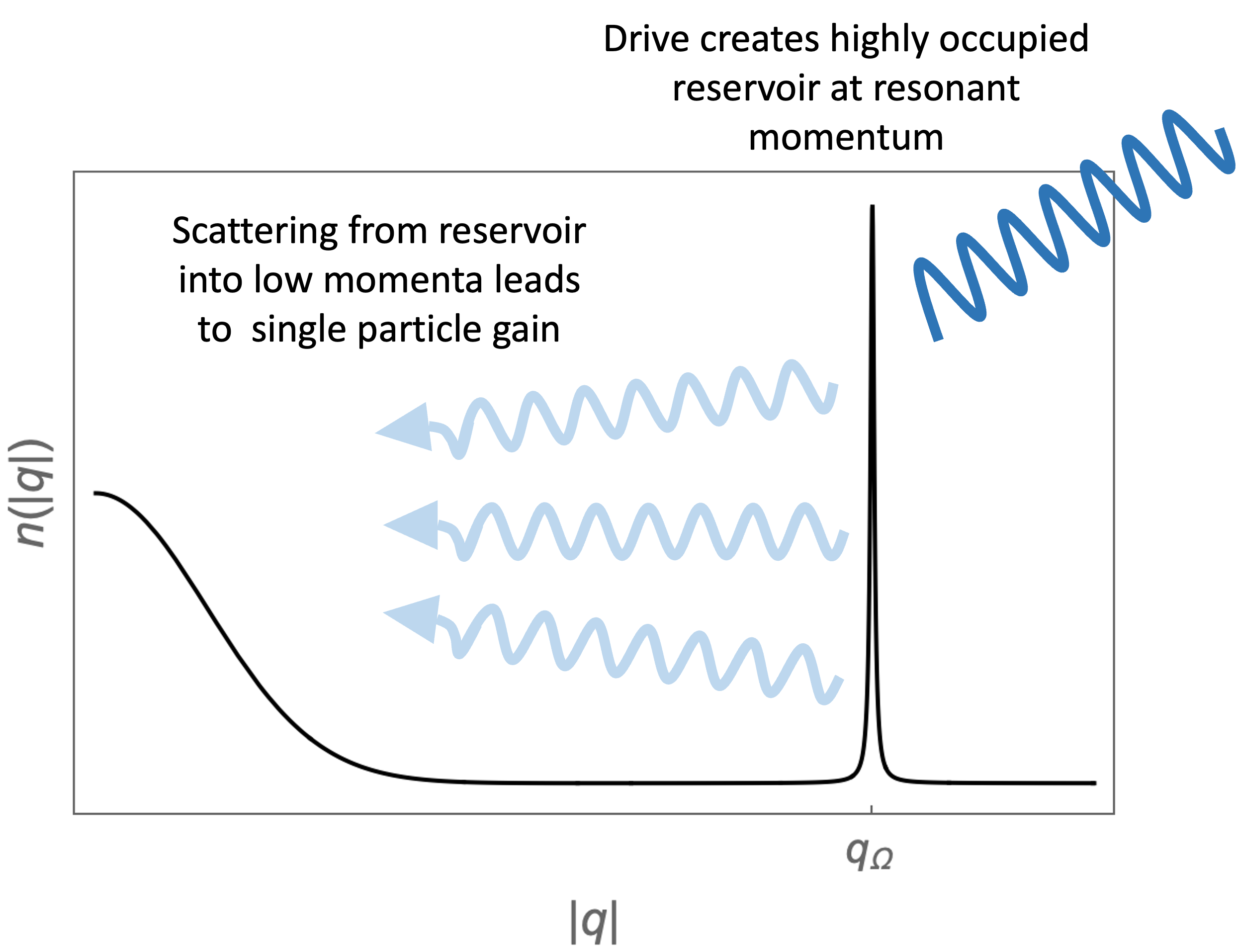}%
}\hfil
    \subfloat[\label{fig:phase_diagram_b}]{%
  \includegraphics[width=0.3\linewidth]{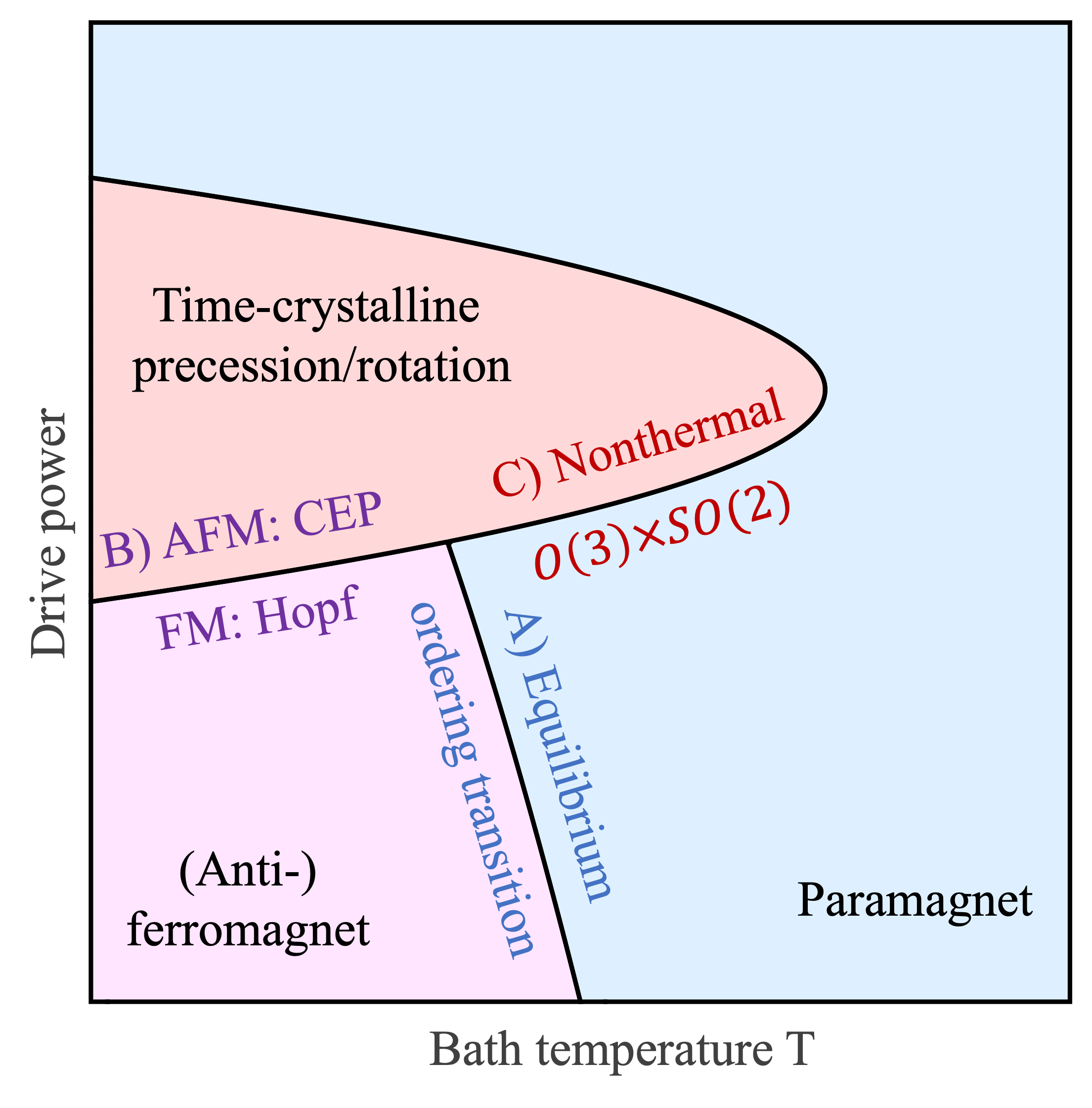}%
}\hfil
    \subfloat[\label{fig:phase_diagram_c}]{%
  \includegraphics[width=0.15\linewidth]{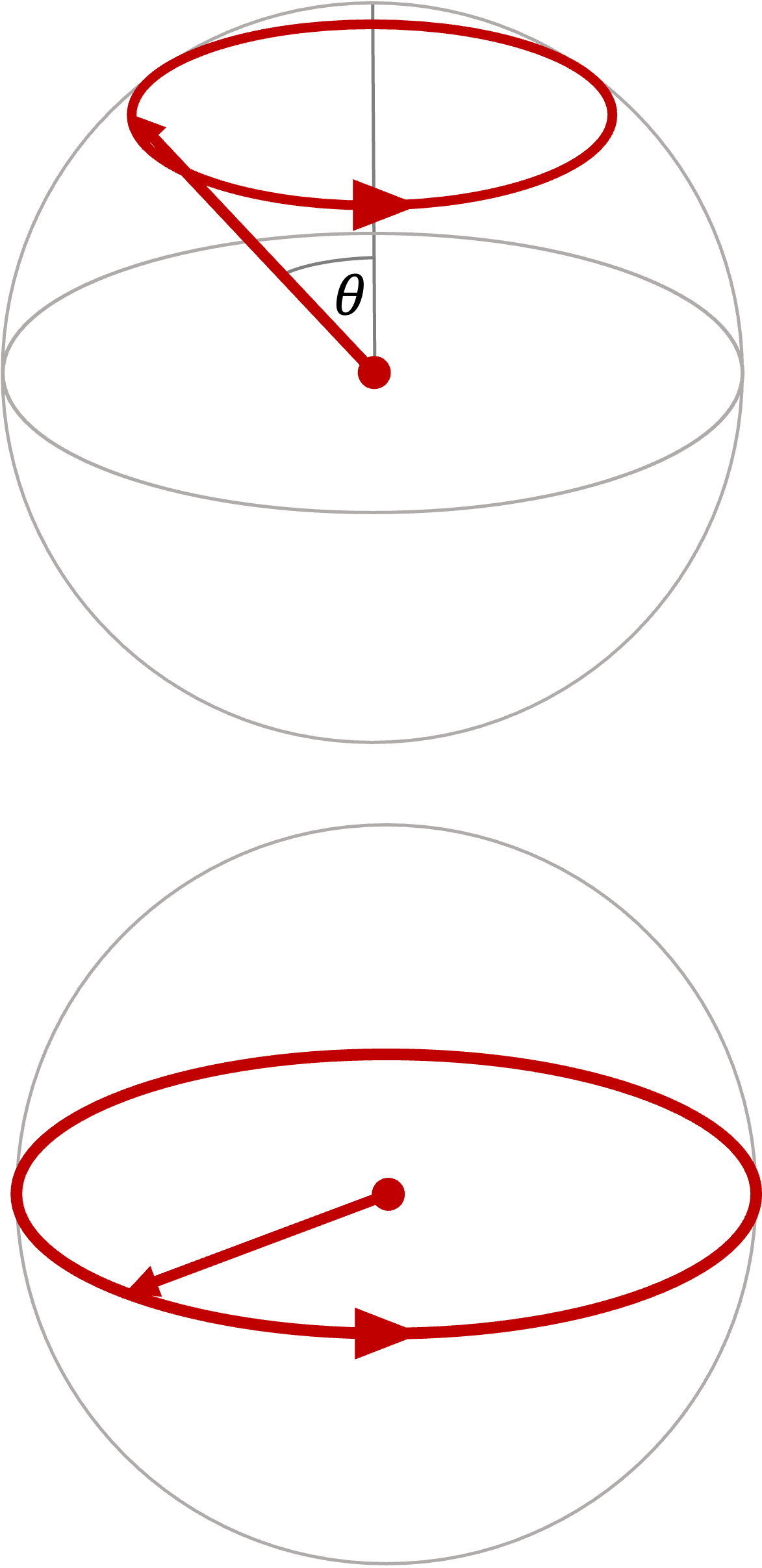}%
}

    \caption{Left panel: Incoherent pump scheme in momentum space. The parametric drive creates a highly occupied reservoir at resonant momentum $q_\Omega$. The reservoir modes incoherently scatter into the long wavelength regime and constitute an effective single particle pump or antidamping. Middle panel: Schematic phase diagram of the driven (anti-)ferromagnet with bath temperature $T$ and drive power proportional to pumping reservoir occupation $\nbath$. At small drives there are the two static phases: the paramagnet and long ranged order antiferro- or ferromagnet. At suitably strong drives and low noise levels, time crystalline order emerges. Too strong drives induce heating and push the system back in a thermal paramagnetic state. Right panel: visualization of time crystalline orders. The order parameter traces out a rotation for the antiferromagnet (AFM) and precession for the ferromagnet (FM). 
    }
    
    \label{fig:phase_diagram}
\end{figure*}
\section{Model} 
Macroscopic orders and the universal features that they share independently of microscopic details in thermal equilibrium are captured by symmetry based field theories. We consider the dynamics of an $N$-component bosonic field $\vecphi\in\mathbb{R}^N$ with a global $O(N)$ symmetry in $d+1$ dimensions. In a closed system it is described by a Lagrangian 
\begin{equation}\label{eq:equilibriumO(N)}
\begin{split}
    \mathcal{L}[\phi]&=-\frac{1}{2}\vecphi^T(\partial_t^2+Z\vec\nabla^2)\vecphi+V[\vecphi],\\
    V[\phi]&=\frac{1}{2}r\vecphi^2+\frac{\lambda}{4}\vecphi^4.
    \end{split}
\end{equation}
This emerges e.g., as an effective theory for magnetic insulators, like the Hubbard model in the strong coupling limit. Here $r\in \mathbb{R}$ controls the equilibrium transition, $Z\in\mathbb{R}$ the excitation dispersion and $\lambda>0$ the scattering rate.\\
To capture dynamical and fluctuation effects, we place this Lagrangian on a Keldysh contour. We linearly couple it to a bosonic Ohmic bath describing the cold thermal reservoir for the system. Integrating out this bath, and performing the Keldysh rotation on the time contour $\vecphi_{c,q}=1/2(\vecphi_+\pm\vecphi_-)$ then yields the following dissipative Keldysh Lagrangian~\cite{kamenev2023,Sieberer2025}:
\begin{equation}\label{eq:Keldysh_Lagrangean}
\begin{split}
    \mathcal{L}_K[\vecphi_{c,q}]=&-\vecphi_q^T(\partial_t^2-Z\vec\nabla^2+2\gamma\partial_t)\vecphi_c+4iD \vecphi_q^2\\
    &-V[\vecphi_c+\vecphi_q]+V[\vecphi_c-\vecphi_q].
    \end{split}
\end{equation}
Here $\gamma$ is the dissipation rate into the thermal bath at  temperature $T=D/\gamma$ (we use units in which $k_B=1$). We have also dropped non-Markovian contributions that are subleading on the large time scales $|t-t'|\gg 1/T$ we are interested in~\cite{kamenev2023}. Note that we cannot drop the second order time derivative term despite the presence of dissipation, since it is necessary to capture the coherent oscillatory effects potentially present in nonequilibrium phases \cite{kaplan2025,zelle2024}. In the presence of a finite thermal noise level $D$, this action can be expanded and truncated in the quantum field $\phi_q$ at the gaussian level. This amounts to a semiclassical limit where quantum fluctuations are dominated by noise fluctuations. This is the generic case for a driven-open system. Reinstating $\hbar$ also shows that $\phi_q$ fluctuations are indeed suppressed by $\hbar$ with respect to $\phi_c$ \cite{kamenev2023}. The resulting semiclassical theory is still fluctuating and interacting and therefore goes way beyond mean field approximations.\\
Such a field theory describes the long wavelength fluctuations of e.g., a Heisenberg antiferromagnet for $N=3$ or a time-dependent Ginzburg-Landau equation of a superconducting order parameter for $N=2$.  Irradiating such systems with a laser manifests as parametric drives in the field theory, i.e. one or more of the couplings become time dependent. Here we focus on parametric oscillations of the mass-like term $r\rightarrow r(t)=r_0+r_D\cos{2\Omega t}$, where the drive period $\Omega^{-1}$ is assumed to be much shorter than the longest time scales of the system $\tau\gg\Omega^{-1}$. On the bare level, this simply leads to parametric resonance for those modes whose eigenfrequency is $\Omega$ \cite{landau1976}. This translates to a resonance condition in momentum space: high momentum modes at $|\vecq| \sim q_\Omega$, where their dispersion is $\omega(q_\Omega) \sim \Omega$ are highly occupied. Therefore, the bare occupation takes the shape 
\begin{align}\label{eq:distribution}
n_0(\vecq)\approx n_\text{b}\delta(|\vecq|-q_\Omega),
\end{align}
where the density of excitations created by the drive $n_\text{b}$ is proportional to the pump power.\\
Before we start to calculate the impact of interactions on this in a pertubation theory, we anticipate their physical role: In the free theory the system cannot redistribute the energy and excitations in frequency and momentum space. This happens through collisions, modes from high frequencies 'rain down' into the low frequency regime. We show that this leads to both an incoherent pumping which manifests as an antidamping and heating, see Fig.~\ref{fig:phase_diagram_a} for a schematic visualization. 

\subsection{Antidamping and heating}

\begin{figure}
    \centering
    \includegraphics[width=0.4\columnwidth]{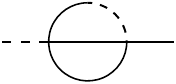}\hfil
    \includegraphics[width=0.4\columnwidth]{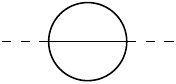}
    \caption{Sunset diagrams that give rise to antidamping (left) and heating (right). Here, solid lines correspond to classical fields and dashed lines to quantum fields in an open system Keldysh path integral. See e.g., \cite{kamenev2023} for a detailed review of diagrammatics in the Keldysh path integral approach.}
    \label{fig:sunsets}
\end{figure}

We now calculate this effect for the low frequency regime perturbatively using real-time Green's functions on the Keldysh contour. As usual, they split into a response function and a correlation function. The bare retarded (advanced) response takes the form 
\begin{align}
    \chi^{R/A}_{ij}(\omega,\vecq)= G^{R/A}_0(\omega,\vecq)\delta_{ij}=\frac{\delta_{ij}}{-\omega^2\mp i\omega\gamma(\vecq)+r(\vecq)},
\end{align}
where $i,j=1,...,N$ for an $O(N)$ symmetric theory. We account for the fact that modes at high momenta can typically dissipate into the bath faster by allowing for a momentum dependent damping. We expand the momentum dependencies of the damping as well as the gap as
\begin{align}\label{eq:momentum_dependence}
    \gamma(\vecq)=\gamma+Z_\gamma\vecq^2,\quad r(\vecq)=r+Z\vecq^2
\end{align}
with $Z,Z_\gamma>0$. These Green's functions are applicable in the low frequency regime, where the fast drive can be averaged out \footnote{This corresponds to a Magnus expansion of Floquet Green's function where we restrict to the lowest order in Floquet space. This is justified in the fast drive limit.}. The correlation function (or Keldysh Green function) encodes the information about the occupation and reads on the bare level:
\begin{equation}
\begin{split}
    &\langle\phi_i(\omega,\vecq)\phi_j(-\omega,-\vecq)\rangle =\mathcal{C}_{ij}(\omega,\vecq)=G^K_0(\omega,\vecq)\delta_{ij},\\ &G^K(\omega,\vecq)=n_b\,\delta(|\vecq|-q_\Omega)\gamma(\vecq)G_0^R(\omega,\vecq)G_0^A(\omega,\vecq).
    \end{split}
\end{equation}
The effective damping is encoded in the frequency dependence of the self energy of the retarded response, $\delta\gamma=i\partial_\omega\Sigma^R(\omega=0,\vecq=0)$. The noise level or effective temperature and thereby the heat up of the system is encoded in the self energy of the Keldysh Green function $\delta D=\Sigma^K(\omega=0,\vecq=0)$. For $\phi^4$ theories the leading contribution to both $\delta\gamma$ and $\delta D$ is second order in perturbation theory as well as in a large $N$ expansion. The technical explanation is, that redistributing modes from high to low frequencies and momenta requires frequency and momentum dependent loop corrections, which only arise through the so called sunset diagrams, see Fig \ref{fig:sunsets}. Therefore, it is crucial to include the second order contributions in a nonequilibrium setting to get the correct long-time steady state behavior. Since there is a finite Markovian noise level we can restrict the Keldysh path integral \eqref{eq:Keldysh_Lagrangean} to the quadratic level in the quantum field $\phi_q$ while keeping higher order linearities in $\phi_c$. The neglected non-Gaussian terms in the quantum field are suppressed by a relative $\hbar$ \cite{kamenev2023}. Then, the relevant corrections to leading order in perturbation theory read
\begin{widetext}
\begin{equation}
\begin{split}
    \Sigma^R(\omega,\vecq=0)=\frac{3N\lambda^2}{8(2\pi)^{2d+2}}\int_{\vecq_1,\vecq_2,\vecq_3}\int_{\omega_1,\omega_2,\omega_3}&\delta(\omega_1+\omega_2+\omega_3+\omega)\delta(\vecq_1+\vecq_2+\vecq_3)\\
    &G^K_0(\omega_1,\vecq_1)G^K_0(\omega_2,\vecq_2)G^R_0(\omega_3,\vecq_3)
    \end{split}
\end{equation}
and
\begin{equation}
\begin{split}
    \Sigma^K(\omega=0,\vecq=0)=\frac{2N\lambda^2}{(2\pi)^{2d+2}}\int_{\vecq_1,\vecq_2,\vecq_3}\int_{\omega_1,\omega_2,\omega_3}&\delta(\omega_1+\omega_2+\omega_3)\delta(\vecq_1+\vecq_2+\vecq_3)\\
    &G^K_0(\omega_1,\vecq_1)G^K_0(\omega_2,\vecq_2)G^K_0(\omega_3,\vecq_3).
    \end{split}
\end{equation}

\end{widetext}
Using the bare Green's function as above, we can carry out the loop integrations explicitly. See appendix \ref{app:bath_eliminations} for a step-by-step calculation. This yields a antidamping of
\begin{align}
    \delta\gamma&=-\mathcal{N}_\gamma \lambda^2 (q_\Omega\,n_\text{b})^2,
\end{align}
 where $\mathcal{N}_{\gamma}>0$ collect model constants and is provided in the appendix and a heating

\begin{align}
    D\rightarrow D+\delta D,\,\delta D=\frac{\lambda^2}{\Omega^2}\mathcal{N}_D (q_\Omega \,n_\text{b})^3,
\end{align}

where the normalization $\mathcal{N}_D>0$ is again given in the appendix. This is the key result of the perturbative self energy calculation: It shows that the parametric drive does not only enter the steady state phase diagram as a shift of the temperature, it can also lead to an antidamping instability. 

\subsection{Nonlinear dampings}

In the 'antidamped' regime nonlinear contributions to the damping that enter the Keldysh Lagrangian as $u(\vecphi_q^T\partial_t\vecphi_c)(\vecphi_c^T\vecphi_c)$ and $u'(\vecphi_q^T\vecphi_c)(\vecphi_c^T\partial_t\vecphi_c)$ become necessary to stabilize the system. Here we show that they are generated in perturbation theory, as well, even if completely absent in the original model. The leading order contributions read
\begin{align}
    u&=i\frac{2\lambda^2}{(2\pi)^{d+1}}\int_{\vecq,\omega}\left(\partial_\omega G_0^K(\omega,\vecq)\right)G_0^R(\omega,\vecq)+\mathcal{O}(\lambda^3),\\
    u'&=i\frac{(N^2+1)\lambda^2}{(2\pi)^{d+1}}\int_{\vecq,\omega}\left(\partial_\omega G_0^K(\omega)\right)G^R_0(\omega,\vecq)+\mathcal{O}(\lambda^3).
\end{align}
The loop integral over the reservoir Green's functions can be performed straightforwardly, in the same manner as in the self energy contributions and we find
\begin{align}
    u&=\frac{\lambda^2(\Omega^2+5\gamma_\Omega^2)q_\Omega^{2}\nbath\delta q}{4\gamma_\Omega(\Omega^2+\gamma_\Omega^2)^2(2\pi)^{3}}>0\\
    u'&=\frac{N^2+1}{2}u.
\end{align}
We thus have established, that the nonlinearities stabilizing the limit cycle phases are genereted with the correct signs by integrating out the pumping reservoir perturbatively. 

\subsection{Phase diagram}

Based on the previous derivations we conclude that the slow degrees of freedom of the rapidly driven model are captured by an effective Keldysh Lagrangian of the form
\begin{equation}
\begin{split}
    \mathcal{L}_{eff}=\vecphi_q^T\Big[&\partial_t^2\vecphi_c+(2\bar\gamma+u\rho_c)\partial_t\vecphi_c+u'(\partial_t\rho_c)\vecphi_c\\
    &+\frac{\delta \bar V[\vecphi_c]}{\delta\vecphi_c}+2i\bar D\vecphi_q\Big],    
\end{split}
\end{equation}
where $\rho=\vecphi^T\vecphi$ and the bars indicate that the drive shifts the effective couplings with respect to their bare values. The major effect of the drive on this effective long wavelength model is that it allows tuning damping. This opens up a new axis in the phase diagram of the $O(N)$ models along which genuine nonthermal orders can emerge, as shown in \cite{zelle2024}.\\
In absence of the drive, the model has a one-dimensional phase diagram controlled by the mass term of the potential $V$. Upon tuning it through zero there is a Model A type equilibrium phase transition \cite{Hohenberg1977}. The normal phase is destabilized and a finite nontrivial order parameter $\vecphi_0\neq 0$ is stabilized by the nonlinear contribution of $V$ spontaneously breaking the $O(N)$ symmetry to $O(N-1)$. This is the famous sombrero  potential. Analogously, if a nonthermal driving effectively tunes the damping rate through zero, all time independent field configurations are destabilized. Instead, limit cycle solutions are stabilized by the nonlinear nonequilibrium forces $u,u'$, that are disallowed at thermal equilibrium. The simplest one is a rotation at fixed amplitude, $\vecphi_0(t)=\vecphi_0(\cos{\omega_0t},\sin{\omega_0t},0,...,0)^T$ with $\vecphi_0=\sqrt{-2\gamma/u}$ and $\omega_0^2=r+\lambda\vecphi_0^2$. Note that the frequency of the limit cycle is set by the system parameters, $\omega_0^2\approx r+\lambda\frac{\gamma-\delta\gamma}{u}$, and not by the drive frequency. Therefore, it can be tuned continuously and is in general incommensurate with and much lower than the drive frequency. Hence, the limit cycle phase constitutes a continuous time crystal. The plane of rotation is picked spontaneously, breaking $O(N)$ to $O(N-2)$. This is stable for $u>u'$ as is the case for parametrically driven models, as we showed above. For $u<u'$ perpetual anharmonic oscillations are stabilized \cite{daviet2024}. The critical theories for the transitions in this model have been worked out in \cite{zelle2024} and \cite{daviet2024}. The transition between a normal phase and the limit cycle order defines a genuinely nonthermal universality class that has a $O(N)\times SO(2)$ symmetry. The transition between the ordered phase of the $O(N)$ and the rotating phase is governed by a so called critical exceptional point which ultimately gives rise to a fluctuation-induced first-order phase transition.\\ 
At this point, we note that the antidamping instability is not triggered in any driven material that abides to a $\phi^4$-like description. Since the heating effect scales as $n_\text{b}^3$ while the antidamping only as $n_{\text{b}}^2$, the time-crystalline phase only emerges if the initial damping $\gamma$ is small, so that thermal noise does not destroy any ordering tendency before the instability is reached. The resulting schematic phase diagram including the universality classes of the transition is depicted in \ref{fig:phase_diagram_b}.
\\
\subsection{The Ferromagnet - $O(3)$ vs $SO(3)$}
We also highlight the important special case of an $SO(3)$ symmetry group (as distinct from $O(3)$ discussed above), describing  isotropic ferromagnets, for example. In this case, an additional symmetry allowed cubic coupling contributes to the Lagrangian, of the form
\begin{align}\label{eq:so(3)_coupling}
    \kappa \vecphi_q\cdot(\partial_t\vecphi_c\times\vecphi_c).
\end{align}
This coupling changes the limit cycle itself from a rotation on a grand circle to a precession of the form
\begin{align}
    \langle\vecphi(t)\rangle=\phi_0\left(\sin{\theta}\cos{\omega_0t},\sin{\theta}\sin{\omega_0t},\cos{\theta}\right)^T,
\end{align}
where the precession angle reaches $\theta\rightarrow \pi/2$, corresponding to a rotation along a circle, as $\kappa\rightarrow 0$. The dynamic limit cycles are visualized in Fig.~\ref{fig:phase_diagram_c}. It furthermore alters the loop structure of the scattering between reservoir and long wavelength modes but does not change the essential property $\delta\gamma<0$. The detailed loop calculations are provided in appendix \ref{app:bath_eliminations}. \\
The additional interactions also impact the universal behavior at the transitions. At the transition between the normal and the rotating phase, the operator distinguishing a $SO(3)\times SO(2)$ and a $O(3)\times SO(2)$ symmetric fixed point is of sixth order in field fluctuations and thereby irrelevant in an RG sense not changing the universality class. The situation is different for the transition between an ordered and precessing phase. The critical exceptional transition of the $O(3)$ model happens at a vanishing onset frequency. This is not true in the presence of the coupling $\kappa$, \eqref{eq:so(3)_coupling}, where the frequency at the transition is set by $\omega_{0}^2=\phi_0^2\kappa$. Therefore, the transition is captured by a Hopf bifurcation, which lies in the $XY$ universality class \cite{Risler2004, Sieberer2013}. We give the detailed mapping in appendix \ref{app:RG}.

\begin{figure*}
    \centering 
    \subfloat[\label{fig:NumRot}]{%
  \includegraphics{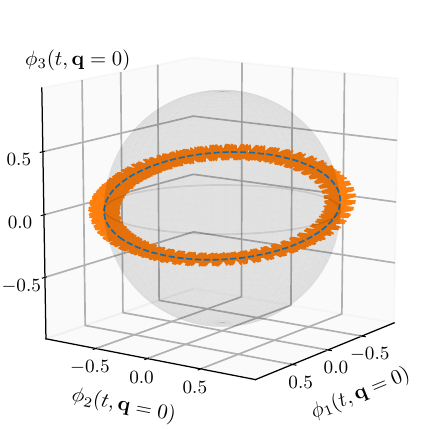}%
}\hfil
    \subfloat[\label{fig:NumFreq}]{%
  \includegraphics{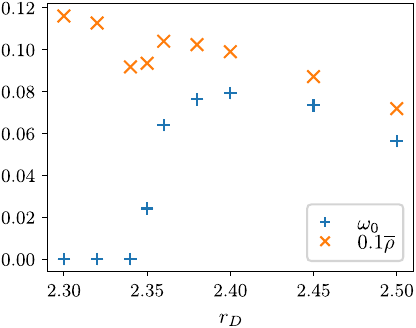}%
}
\caption{Numerical simulations of Eq.~\eqref{eq:EoM} with $-r_0=Z=1, \lambda=0.5, 2\gamma=10^{-3}$ and $T=0$ and full $O(3)$ symmetry, $\kappa=0$. The driving frequency $\Omega=2.15$ is chosen to have a parametric resonance of the longitudinal mode around momentum $q_\Omega \sim \pi/2$. (a) The long time solution with $r_D=2.4$ is given by the orange solid line. The blue dashed line shows the same trajectories averaged over the fast amplitude oscillations (with frequency $\Omega$). The order parameter indeed traces out a circle. (b) Frequency of the limit cycle $\omega_0$ (blue) and its amplitude $\overline{\rho}$ (orange) averaged over the fast oscillations as a function of the driving power $r_D$. We see that at the transition from ordered to rotating phase, the amplitude dips, which is in line with the critical exceptional point scenario for the transition (see main text).}
    \label{fig:Numerics}
    \end{figure*}
\section{Numerical simulations} 

Our analytical derivations show that the parametrically pumped, dissipative $O(N)$ model can host a stable nonequilibrium state in the form of a slow limit cycle phase. We now corroborate these results by direct numerical simulations. As laid out above, fluctuations of the quantum field $\phi_q$ in the Keldysh action of our model \eqref{eq:Keldysh_Lagrangean} can be savely treated at a quadratic level due to the finite noise level $D$ \cite{kamenev2023} while keeping the full nonlinearity in the classical field. Then, the path integral is equivalent to the Langevin equation
\begin{equation}\label{eq:EoM}
    \Big(\partial_t^2+(2\gamma-Z_1\nabla^2)\partial_t+r(t)-Z\nabla^2+\lambda|\vecphi|^2\Big)\vecphi+\xi=0.
\end{equation}
Here, $\xi$ is a Gaussian white noise whose width is given by $\langle \xi_i(t,\xx),\xi_j(t',\xx')\rangle=D\delta(t-t')\delta(\xx-\xx')$ with an explicitly time dependent $r(t)$. Mimicking an antiferromagnet, we perform these simulations at $r<0$, as explained above. The spin-wave velocity is set to unity and we assume sufficiently weak dampings. Simulations on a cubic lattice of length $L=32$ show that indeed initially a sufficient drive induces a high occupation of longitudinal modes at momenta $q_\Omega$. These quickly scatter into the low momentum/frequency regime and lead to a macroscopic occupation of transversal modes at $\vecq=0$, and at a finite frequency $\omega_0$. As predicted theoretically, there is a coherent, long-range ordered rotation of the order parameter, visualized in Fig.~\ref{fig:NumRot}. Along the transition between antiferromagnet and time crystalline order, the mapping to the critical exceptional point of driven $O(N)$ models predicts a decrease of the amplitude at the transition due to exceptionally enhanced fluctuations and a square root behavior of the angular velocity upon entering the rotating phase~\cite{zelle2024}. Both are confirmed by numerical simulations as shown in Fig.~\ref{fig:Numerics}.
\\
\section{Pumped magnetic materials} 
Let us now comment on possible realizations of this nonthermal order in physical setups. An important physical example, where our mechanism could be tested, is an isotropic antiferromagnetic insulator, whose long wavelength dynamics are naturally described by $O(N)$ models~\cite{Sachdev2011,Fradkin2013}. This points, for instance, at certain cuprates \cite{Kastner98,Headings2010,Guarise2010} or perovskite antiferromagnets \cite{Coldea98,Yamaguchi99}.\\
The closed system ($\gamma=0$) then has two gapless spin-wave excitations transversal to the ordering vector with ballistic dispersions $\omega_T(\vecq=0)=\pm v|\vecq|$ and velocity $v=\sqrt{Z}$. The longitudinal fluctuations have a frequency gap of  $\Delta_L=2|r|\propto 2J$, where $J$ is the spin exchange coupling. The remaining interaction parameter $\lambda$ sets the scattering length of magnetic excitations.
\\
In practice, the magnetic system is not closed and phononic bathes generically give rise to thermal noise fluctuations and dissipation at a fixed temperature, so that neither energy nor total magnetization are conserved. These can be modeled by dissipation and noise rates $\gamma(\vecq)$ and $D(\vecq)$ with a potential momentum dependence that still obeys the fluctuation-dissipation relations of thermal equilibrium, $\gamma(\vecq)/D(\vecq)=T=\text{const.}$ They typically increase monotonically with momentum $\vecq^2$, since there is more phase space for scattering with bath degrees of freedom. In our setup, this is captured by $Z_\gamma$ in Eq.~\eqref{eq:momentum_dependence}. The zero momentum mode of the dissipation, $\gamma(\vecq=0)$, gives the decay rate of the total $O(N)$ charge (e.g., magnetization) of the system, while energy is dissipated at all momenta. The dissipation at the reservoir-momentum $\gamma(\vecq_\Omega)\equiv\gamma_\Omega$ gives a direct dissipation from the reservoir into the bath and does not destabilize the proposed mechanism. Parametric driving can, for instance, emerge through oscillating electric fields. Numerical simulations of Hubbard models at strong coupling and half filling that are driven by electric fields confirm this and show an antidamping instability at low frequencies due to the rapid parameteric drive \cite{Walldorf19,okugawa2026}.\\
Let us consider the classical Heisenberg magnet $\text{KNiF}_3$ as an example, where $J\approx8\,\text{meV}$ \cite{Yamaguchi99, Martin1997}. The parameters of the numerics reported in Fig.~\ref{fig:Numerics} correspond to a drive frequency of $4.2\,\text{THz}$ and assume a damping of $2\gamma\approx 2\,\text{GHz}$. The most direct evidence of the rotating phase is the light emitted by the slowly rotating macroscopic magnetic moment. This should further appear in magnetic resonance. The critical scaling behavior at the transition \cite{zelle2024,daviet2024} e.g. manifests itself in the singularities of the dynamical spin structure factor.

\section{Discussion and Outlook} In this paper, we have presented a universal mechanism to capture nonequilibrium condensation phenomena within parametrically driven, open field theories. We show that rapid parametric drives do not only tune the parameters of an effective free energy landscape but enter as a nonequilibrium axis in the phase diagram along which ordered, nonthermal steady state phases arise. These are genuine nonequilibrium steady states that cannot be captured by an effective free energy. Technically, it provides a method to model pumping effects stemming from rapidly oscillating drives effectively by noisy, non-Hermitian field theories that feature an effective time translation invariance on the 'mesoscopic' scale of description, but manifestly break equilibrium conditions. In these effective theories, the transition is triggered by tuning an effective damping rate through zero. Although our analysis is based on an $O(N)$ symmetric Lagrangian, the mechanism is more general and also applies to other symmetry groups. Such a symmetry analysis and the ensuing effective field theory should also describe the universal behavior of magnon condensation in parametrically driven anisotropic magnets \cite{Rezende2009,Demokritov2006,NowikBoltyk2012}.\\
The macroscopic limit cycle characterizing the nonequilibrium phases spontaneously breaks the effective time translation symmetry of the $O(N)$ model, and thus constitutes a dissipative time crystal. Our theory indicates that such phases can be realized by simple driving schemes in magnetic materials if the damping rate of the undriven system is small. This complements recent approaches to generate limit cycle phases in magnetic systems with lower symmetry \cite{zelle2024}. It furthermore provides a promising novel pathway for devising the effective field theories of light-induced orders. \\
Such limit cycle orders have drawn tremendous attention in active matter scenarios, in particular in the context of nonreciprocal interactions \cite{Fruchart2021,Hanai2019,Martin2025, pisegna2024, brauns2024, shankar2022, dinelli2023,begg2024}. These can induce "hunt and flight" dynamics that stabilize limit cycles. Much effort has been devoted to the realization of nonreciprocal interactions on microscopic scales, particularly in spin systems \cite{Hanai2025, nadolny2025}. This work shows that these active phases can emerge in driven materials as well.\\
Further, scaling laws within the time-crystalline phase can be expected to qualitatively deviate from equilibrium phenomenology. On the one hand, in analogy to driven-dissipative exciton-polariton condensation, there will be a mode with a Kardar-Parisi-Zhang (KPZ) nonlinearity, which leads to subdiffusive transport \cite{daviet2025}. It will, however, interact with the soft modes from the broken internal symmetries, paving the way to new nonthermal fixed points in the respective nonlinear sigma models. Additionally, recent work on similar phases in $O(2)$ ferrimagnets in one dimension has shown that these can constitute 'active magnets' with self propelled defects, whose relaxational dynamics differs starkly from equilibrium physics~\cite{hardt2025}.
\\

\begin{acknowledgments}

 We thank A. Chakraborty, R. Hanai, M. Kalthoff, J. Lang, A. Rosch and R. Tazai for useful discussions. We acknowledge support by the Deutsche Forschungsgemeinschaft (DFG, German Research Foundation) CRC 1238 project C04 number 277146847. CZ was supported by the DFG through project number 570906600. 
\end{acknowledgments}

\bibliography{research_statement}
\clearpage

\onecolumngrid
\setcounter{equation}{0}
\setcounter{figure}{0}
\setcounter{table}{0}
\setcounter{page}{1}
\makeatletter
\renewcommand{\theequation}{S\arabic{equation}}
\renewcommand{\thefigure}{S\arabic{figure}}
\renewcommand{\thetable}{S\arabic{table}}
\renewcommand{\bibnumfmt}[1]{[S#1]}
\renewcommand{\citenumfont}[1]{S#1}
\acresetall

\appendix

\section{ Perturbative elimination of rapid drive}\label{app:bath_eliminations}

In this appendix we give the details on the elimination of the highly occupied reservoir modes at large momenta $q_\Omega$.To that end, we assume that the occupation of modes is peaked at these momenta, so that we can write the occupation function as $n(q)\approx\delta(q-q_\Omega) \cdot  n_b$, where $n_b$ is the number of modes that the pump injects at the resonent momentum $q_\Omega$ and increases with pump power. This effect already occurs on the level of the free theory.\\
Our goal is to derive the corrections to the response as well as occupation functions of the low frequency regime that arise by the scattering of these modes into the low frequency regime. We will do this in a perturbative diagrammatic approach using nonequilibrium Green functions, or equivalently a perturbative expansion of the MSRJD path integral \cite{Taeuber2014}.\\
For our model, \eqref{eq:EoM}, the bare retarded and advanced responses have the form
\begin{align}
    \chi^{R/A}_{ij}(\omega,\vecq)= G^{R/A}_0(\omega,\vecq)\delta_{ij}=\frac{\delta_{ij}}{-\omega^2\pm i\omega\gamma(\vecq)+r(\vecq)},
\end{align}
where $i,j=1,...,N$ for an $O(N)$ symmetric theory and we expand the momentum dependencies of the damping as well as the gap as
\begin{align}
    \gamma(\vecq)=\gamma+Z_\gamma\vecq^2,\quad r(\vecq)=r+Z\vecq^2
\end{align}
with $Z_\gamma,Z>0$. 
The correlation function is
\begin{align}
    \langle\phi_i(\omega,\vecq)\phi_j(-\omega,-\vecq)\rangle =\mathcal{C}_{ij}(\omega,\vecq)=G^K_0(\omega,\vecq)\delta_{ij},\quad G^K(\omega,\vecq)=:f(\omega,\vecq)G_0^R(\omega,\vecq)G_0^A(\omega,\vecq).
\end{align}
Using the occupation of the reservoir modes, we can fix the bare correlation function $G^K_0$ of the reservoir 
\begin{align}
    n(\vecq)=\int_{\omega}G^K_0(\omega,\vecq)
\end{align}
to wit

\begin{align}\label{eq:gk_bath}
    f(\omega,\vecq)=n_b\,\delta(|\vecq|-q_\Omega)\gamma(\vecq).
\end{align}
The antidamping impact on the slow modes  modes is given by the frequency dependence of the self-energy contributions $\Sigma(\omega)$ . The shift of the damping is encoded in the corrections to the retarded response, i.e., $\delta\gamma=\operatorname{Im}\partial_\omega\Sigma^R_p(\omega=0,\vecq=0)$. The heating effect is encoded in the correction to the inverse correlation $\Delta D=\Sigma^K_p(\omega=0,\vecq=0)$. Below, we derive these corrections perturbatively. 

\paragraph{Antiferromagnet --}
In the case of an antiferromagnet, where $\kappa=0$ and the scattering vertex is the quartic $\lambda\phi^4$ coupling. The leading order contribution to the frequency dependence of $\Sigma^R$ is the two-loop sunset diagram stemming from the $\lambda\phi^4$ coupling. The loop integral reads
\begin{align}
\nonumber
    I_{sunset}(\omega,\vecq=0)=(2\pi)^{-2d-2}\int_{\vecq_1,\vecq_2,\vecq_3}\int_{\omega_1,\omega_2,\omega_3}&\delta(\omega_1+\omega_2+\omega_3+\omega)\delta(\vecq_1+\vecq_2+\vecq_3)\\
    &G^K_0(\omega_1,\vecq_1)G^K_0(\omega_2,\vecq_2)G^R_0(\omega_3,\vecq_3).
\end{align}
Using a Fourier representation of the $\delta$ distributions, we rewrite
\begin{align}
    I_{sunset}(\omega,0)=&(2\pi)^{-3d-3}\int_{\boldsymbol{s},t}e^{it\omega}\left( \int_{\vecq_1,\omega_1}e^{i\boldsymbol{s}\cdot\vecq_1+it\omega_1}G^K_0(\omega_1,\vecq_1)\right)\\
    &\left(\int_{\vecq_2,\omega_2}e^{i\boldsymbol{s}\cdot\vecq_2+it\omega_2}G^K_0(\omega_2,\vecq_2)\right)\left(\int_{\vecq_3,\omega_3}e^{i\boldsymbol{s}\cdot\vecq_3+it\omega_3}G^R_0(\omega_3,\vecq_3)\right).
\end{align}
Now we use rotational invariance of the Greens functions, i.e., that they only depend on $q\equiv|\vecq|$ and restrict ourselves to $d=3$ to perform the angular momentum integrations. To this end we use
\begin{align}
    \int \dd^3\vecq\, e^{i\boldsymbol{s}\cdot\vecq}F(q)= 2\pi\int_{-\infty}^{\infty}q^2\dd q \frac{\sin{qs}}{qs} F(q),
\end{align}
where $s=|\boldsymbol{s}|$. We can now straightforwardly perform the integration over $\boldsymbol{s}$ by $\int \dd^3 \boldsymbol{s} s^{-3}\sin^3(sq_\Omega)=\pi^2$. We then compute the subintegrals $I_1=\int \dd \omega \int q\dd q e^{it\omega}G^R_0(\omega,q)$ and $I_2=\int \dd \omega \int q\dd q e^{it\omega}G^K_0(\omega,q)$. 
Since $G^R$ has  poles in the lower complex  half plane only, $I_1$ is proportional to $\theta(-t)$. Performing the frequency as well as momentum integration yields
\begin{align}
    I_1=\frac{4e^{\gamma t}\pi^2\arctan{Z/Z_\gamma}}{Z}\theta(-t).
\end{align}
For $I_2$ we can use the momentum shell constraint of \eqref{eq:gk_bath} to wit, for $t<0$
\begin{align}
    I_2=4\pi q_\Omega \delta q \nbath\int_\omega e^{it\omega} \frac{2i\gamma_\Omega}{(\gamma_\Omega^2+(\omega-\Omega)^2)(\gamma_\Omega^2+(\omega+\Omega)^2)}=-\frac{4\pi q_\Omega \nbath e^{\gamma_\Omega t}(\gamma_\Omega\sin{\Omega t}-\Omega\cos{\Omega t})}{\Omega},
\end{align}
where $\gamma_\Omega$ is the decay rate into the bath of excitations at $q_\Omega$, $\gamma_\Omega=\gamma(\vecq_\Omega)$. In three dimensions, we thus arrive at
\begin{align}
    I_{sunset}(\omega,0)=2^{-6}\pi^{-3}\int_{-\infty}^\infty \dd t e^{i\omega t}I_1I_2^2.
\end{align}
We are only interested in the frequency derivative at $\omega=0$:
\begin{align}
    \partial_{\omega}I_{\mathrm{sunset}}(\omega=0,\vecq=0)=2^{-6}\pi^{-3}\int_{-\infty}^\infty \dd t\, i t I_1I_2^2.
\end{align}
This integral can be performed analytically. For $\Omega\gg\gamma_\Omega\gg\gamma$ it yields
\begin{align}
    \partial_{\omega}I_{\mathrm{sunset}}(\omega=0,\vecq=0)\approx -i \frac{(q_\Omega \,\nbath)^2\arctan{Z/Z_\gamma}}{256Z\gamma_\Omega^2}
\end{align}
 The full self energy contains additional factors of $(\lambda/4)^2$, the bare scattering of the original theory, a combinatorial factor of $2\cdot3$ and a factor $N$ from the traces over the internal $O(N)$ indices and thus
\begin{align}
    \partial_\omega\Sigma_p^R(\omega=0,\vecq=0)=\frac{3N\lambda^2}{8}\partial_{\omega}I_{\mathrm{sunset}}(\omega=0,\vecq=0).
\end{align}
Therefore, the shift of the damping for the antiferromagnet ($N=3$) to leading order in perturbation theory is
\begin{align} 
    \delta\gamma=\operatorname{Im}{\partial_\omega\Sigma_p^R(\omega=0,\vecq=0)}\approx-\mathcal{N}_\gamma \frac{(q_\Omega\,\nbath)^2}{\gamma_\Omega^2}<0,
\end{align}
proving that the pumping effectively leads to an antidamping contribution. Here $\mathcal{N}_\gamma=\frac{\arctan{Z/Z_\gamma}}{256 Z}$.\\

Equivalently, the pumped bath will also lead to a noise increase $\delta D$ in the low energy modes that will counteract ordering instabilities. This also manifests in a self energy contribution stemming from a sunset diagram with three internal Keldysh Green functions,
\begin{align}
    \delta D=N\lambda^2 2^{-5}\pi^{-3}\int_{-\infty}^{0}\dd t I_2^3\approx\frac{5(q_\Omega\,\nbath)^3}{32\Omega^2},
\end{align}
where we symmetrised the $t$ integration which initially runs from $-\infty$ to $\infty$ and again used $\gamma_\Omega\ll\Omega$. 

\paragraph{Ferromagnet}

In the case of a ferromagnet, there is an interaction $\kappa\epsilon_{ijk}\vecphi_{q,i}\partial_t\vecphi_{c,j}\vecphi_{c,k}$ and thus the leading contribution to $\partial_\omega\Sigma^R_p$ stems from a one loop integral. Here, care needs to be taken when contracting the antisymmetric vertices. The configuration in which the time derivative of the vertex hits an external leg leads to a contribution to the damping
\begin{align}
    \kappa^2\epsilon_{ijk}\frac{1}{(2\pi)^4}(\epsilon_{jik}+\epsilon_{kij})\int_{\omega,\vecq}(-i\omega)G^R_0(\vecq,\omega)G_0^K(\vecq,\omega)=0.
\end{align}
The remaining contributions are
\begin{align}
\nonumber
   \partial_\omega\Sigma^R_p= &\frac{\kappa^2}{(2\pi)^4}\partial_\omega\Bigg|_{\omega=0}\epsilon_{ijk}\Big(\epsilon_{kji} \int_{\nu,\vecq}(-i(\nu-\omega))^2G_0^K(\nu-\omega,\vecq) G^R_0(\nu,\vecq)\\ \nonumber
   &+\epsilon_{jki}(-i(\nu-\omega))G_0^K(\nu-\omega,\vecq) (-i\nu)G^R_0(\nu,\vecq)\Big)\\
        =&-\kappa^2\int_{\nu,\vecq}\nu G^K_0(\nu,\vecq)G^R_0(\nu,\vecq)=-i\kappa^2\frac{\nbath  q_\Omega^2\delta q}{4\pi^3\gamma_\Omega}.
\end{align}
As above, we have established, that the slow spin waves experience an effective antidamping contribution due to the reservoir modes.\\
Similarly to the case of the antiferromagnet, there is also a self energy contribution to the noise, which reads

\begin{align}
    \delta D=2\kappa^2\int_{\omega,\vecq}G^K_0(\omega\vecq)^2\approx  \frac{(\nbath  q_\Omega^2\delta q)^2}{2\pi \gamma_\Omega}.
\end{align}

\subsection{Nonlinear dampings}
The nonlinear damping contributions to the Keldysh Lagrangian $u(\vecphi_q\cdot\partial_t\vecphi_c)(\vecphi_c\cdot\vecphi_c)$ and $u'/2(\vecphi_q\cdot\vecphi_c)\partial_t(\vecphi_c\cdot\vecphi_c)$ are crucial to stabilize the limit cycle phases. Since they are RG relevant in the vicinity of the transition into the rotating and oscillating phases below four dimensions, as the prior analysis showed \cite{zelle2024,daviet2024}, we expect them to be generated as soon as symmetries allow them upon coarse graining close to the antidamping instability. Here we show explicitly how they are generated on the one-loop level as above. \\
To that end we first look at the full frequency dependence, at vanishing external momenta, of the vertex in leading order perturbation theory
\begin{equation}
\begin{split}
\Gamma^{(13)}(\omega_1,\omega_2,\omega_3)=&\frac{\delta^4\Gamma}{\delta\vecphi_{q,a}(-\sum\omega_i)\vecphi_{c,a}(\omega_1)\vecphi_{c,b}(\omega_2)\vecphi_{c,b}(\omega_3)}=\lambda-\lambda^2(2\pi)^{-d-1}\int_{\vecq,\omega}\Big(N^2 G^K_0(\omega-\omega_2-\omega_3,\vecq)G^R_0(\omega,\vecq)\\
&+G^K_0(\omega-\omega_1-\omega_3,\vecq)G^R_0(\omega,\vecq)+G^K_0(\omega-\omega_1-\omega_2,\vecq)G^R_0(\omega,\vecq)\Big)+\mathcal{O}(\lambda^3)
\end{split}
\end{equation}
Now, we identify 
\begin{align}
    u&=i\partial_{\omega_1}\Gamma^{(13)}(\omega_1,\omega_2,\omega_3)\Big|_{\omega_i=0},\\
    u'&=\frac{i}{2}\left(\partial_{\omega_2}+\partial_{\omega_3}\right)\Gamma^{(13)}(\omega_1,\omega_2,\omega_3)\Big|_{\omega_i=0},
\end{align}
to find that
\begin{align}
    u&=i\frac{2\lambda^2}{(2\pi)^{d+1}}\int_{\vecq,\omega}\left(\partial_\omega G_0^K(\omega,\vecq)\right)G_0^R(\omega,\vecq)+\mathcal{O}(\lambda^3),\\
    u'&=i\frac{(N^2+1)\lambda^2}{(2\pi)^{d+1}}\int_{\vecq,\omega}\left(\partial_\omega G_0^K(\omega)\right)G^R_0(\omega,\vecq)+\mathcal{O}(\lambda^3).
\end{align}
The loop integral over the reservoir Green's functions can be performed straightforwardly, in the same manner as in the self energy contributions and we find
\begin{align}
    u&=\frac{\lambda^2(\Omega^2+5\gamma_\Omega^2)q_\Omega^{2}\nbath\delta q}{4\gamma_\Omega(\Omega^2+\gamma_\Omega^2)^2(2\pi)^{3}}>0.\\
    u'&=\frac{N^2+1}{2}u
\end{align}
We thus have established, that the nonlinearities stabilizing the limit cycle phases are genereted with the correct signs by integrating out the pumping reservoir. Furthermore, we showed that $u'>u$ in leading order perturbation theory and the system will go into the rotating phase rather than showing amplitude oscillations \cite{zelle2024}.\\

\section{2. Universal scaling in the presence of ferromagnetic coupling.}\label{app:RG}

Here we consider the case of an $SO(3)$ but not $O(3)$ symmetric theory. This is e.g., a ferromagnet in contrast to a Heisenberg antiferromagnet, where the $\mathbb{Z}_2$ redundancy of the Neél vector causes the system to be fully $O(3) = \mathbb{Z}_2\ltimes SO(3)$ symmetric. The coupling distinguishing those two cases is the precessing force term $\sim \kappa$, see Eq. \eqref{eq:so(3)_coupling}. For larger $N$, the difference between $SO(N)$ and $O(N)$ is encoded in higher order couplings which are irrelevant at the transitions. 
\\
For only $SO(3)$ symmetry, the transition between paramagnet and time crystal is described by the nonthermal $O(3)\times SO(2)$ universality class recently found in \cite{daviet2024} . The transition between ferromagnet and time crystal is instead described by the universality class of the noisy Hopf bifurcation \cite{Taeuber2014,Risler2004}, as we will demonstrate now.

\subsection{2.1 Ferromagnet to time crystal}
We start with the ordered phase in the presence of the ferromagnetic coupling $\kappa$. Without loss of generality, we assume the magnetic order to be oriented along the $z$ axis and expand fluctuations around the mean-field solution $\phi=(m_1,m_2,\sigma+\sigma_0)$. The longitudinal fluctuations $\sigma_0$ are gapped and we restrict ourselves to the dynamics of the Goldstone modes $m_{1,2}$. It is useful to collect the two Goldstone modes into one complex field $m=m_1+im_2$ and switch to a phase-space description of dynamics by introducing the corresponding conjugate field $\pi=\partial_t m$. Linearizing the field theory around the static order in this basis yields 
\begin{align}\label{eq:hopf}
    &\partial_t\pi +(\delta+i\sigma_0\kappa-Z\nabla^2)\pi - Z\nabla^2 m +\xi=0,\\
    &\partial_t m= \pi.
\end{align}
Since $m$ is a Goldstone mode it has to gapless. This implies that $m$ can only appear with derivatives acting on it. Further, the dynamics has to be invariant under the unbroken subgroup $U(1)$. Here, $\delta=2\gamma+u\sigma_0^2$ is the tuning parameter that triggers the transition between static ($\delta>0$) and rotating ($\delta<0$) phase, which spontaneously breaks the $U(1)$ group. Eq. ~\eqref{eq:hopf} reveals that the universal fluctuations are captured by a noisy Hopf bifurcation of the conjugate momentum $\pi$ coupled to the gapless Goldstone modes $\operatorname{Re}m,\operatorname{Im}m$. The finite frequency $\kappa\sigma_0\neq0$ at which the Hopf bifurcation takes place is induced by $\kappa\neq 0$ and prevents the transition from occurring through a critical exceptional point, which would render it first order. This is a consistent mean field picture of a precession, at the transition we see a finite frequency condensation of the conjugate spin wave momentum $\vec{\pi}=\pi_0(\cos{\sqrt{\sigma_0\kappa t}},\sin{\sqrt{\sigma_0\kappa t}},0)^T$. Plugging this back into $\vec\phi=(m_1,m_2,\sigma+\sigma_0),\,\partial_t\vec\phi=\vec\pi$ yields exactly the precessing phase from \ref{eq:EoM}.\\
The decoupling of the critical mode $\pi$ from the soft diffusive Goldstone mode $m$ occurs through their frequency separation. To study the universal fluctuations of the critical mode, we transform into its rotating frame, $\pi\rightarrow e^{iEt}\pi,\,m\rightarrow e^{iEt}m$ with $E=\sigma_0\kappa$. This gives a mass-like contribution to the mode $m$ and, it thus can be integrated out adiabatically, since it moves much faster than the critical mode. This exactly reproduces a noisy Gross-Pitaevskii equation for $\pi$ in the rotating frame,
\begin{align}
        \partial_t\pi+\delta\cdot\pi+(Z_1-iZ_2/E)\nabla^2\pi+\xi_\pi=0.
\end{align}
Including the relevant, symmetry-allowed interactions reproduces a complex Gross-Pitaevskii equation (cGPE). This shows that the universal exponents of the transition between ferromagnet and time-crystal are described by the Hopf or cGPE universality class \cite{Risler2004,Tauber2013a,Sieberer2013}.
\subsection{2.2 Paramagnet to Time crystal}
We now consider the transtion between paramagnet and time crystal, characterised by $SO(3)\times SO(2)$ symmetry breaking. The field theory describing $O(3)\times SO(2)$ symmetry breaking into a time crystalline phase was discussed in \cite{daviet2024}. We thus have to check if the terms distinguishing $SO(3)\times SO(2)$ from $O(3)\times SO(2)$ are contributing to the perturbative RG flow. The relevant degrees of freedom are the ampitude vectors $\vec{\chi}_{1,2}.$ In index notation we write $\chi^i_\alpha$ where upper Roman indices $i=1,2,3$ are the $SO(3)$ indices while the lower Greek indices $\alpha=1,2$ are the $SO(2)$ ones. The simplemost $SO(3)\times SO(2)$ invariants that are \emph{not} also invariant under $O(3)\times SO(2)$ are of the form
\begin{align}
    \left(\epsilon_{ijk}\tilde{\chi}_i^\alpha\delta^{\alpha\beta}\chi^\beta_j\chi^\gamma_k\right)\delta^{\gamma\gamma'}\left(\epsilon_{mno}\left(\nabla^2\chi_m^{\alpha'}\right)\delta^{\alpha'\beta'}\chi^{\beta'}_m\chi^{\gamma'}_n\right)
\end{align}
and permutations thereof. These are thus of second order in derivatives and sextic order in fields and therefore highly irrelevant. Hence, they will not impact the universal exponents in a perturbative expansion and the universal exponents coincide with those of $O(3)\times SO(2)$.

\end{document}